\documentclass[11pt]{article}

\usepackage{amsmath,amssymb}
\usepackage{geometry}
\usepackage{microtype}
\usepackage{cite}
\usepackage[hidelinks]{hyperref}
\title{The MS Unidirectional Current as a Generalization of the ABC Absorption Current}

\author{
Avi Marchewka\\[0.4em]
\small 8 Galei Tchelet Street, Herzliya, Israel\\
\small\href{mailto:Avi.marchewka@gmail.com}{Avi.marchewka@gmail.com}}

\date{}

\begin{document}
\maketitle

\begin{abstract}
For a free scalar particle in one spatial dimension, in a single-pass
half-line geometry, we compare the absorbing Robin boundary condition with the
generalized Marchewka--Schuss (MS) unidirectional-current family. Robin
absorption is characterized by a single parameter $\kappa>0$: once $\kappa$ is
specified, its momentum-dependent absorption profile is fixed. In the
generalized MS construction, by contrast, the detector response is described
by a spectral function $\lambda(k)$, subject to the positivity and
subnormalization conditions of the detection law. We show that every Robin choice
of $\kappa$ corresponds to the particular MS calibration
$\lambda_\kappa(k)=\pi\kappa/(k+\kappa)^2$, which reproduces the complete Robin
absorption current for every admissible one-sided spectral amplitude and at
every time. Thus the entire one-parameter Robin absorption family is contained
within the generalized MS family. The MS construction is more general: the
choice $\lambda_{\rm full}(k)=\pi/(4k)$ gives unit absorption efficiency for
every wave number, which cannot be achieved by any fixed Robin parameter. In
this precise sense, the generalized MS detector process extends the Robin
absorbing-boundary family. We also compare their asymptotic behavior in the
ballistic fixed-ray regime and in the fixed-distance late-time limit.
\end{abstract}

\section{Introduction}

Quantum time-of-arrival theories seek to describe the probability distribution
of the time at which a particle is detected. One broad strategy is to represent
the detection process explicitly as part of the dynamics. The
Marchewka--Schuss construction
\cite{MarchewkaSchuss1998,MarchewkaSchuss2000,MarchewkaSchuss2002} belongs to
this detector-based class. The absorbing-boundary approach has its roots in
Werner's contraction-semigroup formulation of arrival-time observables
\cite{Werner1987} and is realized in the modern ABC model through an absorbing
Robin boundary condition \cite{Tumulka2022,Tumulka2023}. In both MS and ABC,
the detector is therefore part of the theoretical construction rather than an
external interpretation applied after the unitary evolution.

This detector-based strategy is conceptually distinct from prescriptions
that assign arrival-time statistics without specifying an explicit detector
dynamics. Kijowski's covariant construction \cite{Kijowski1974} defines an
arrival-time POVM directly. Another widely used reference, particularly in
scattering and time-of-flight problems, is the free Schr\"odinger probability
current and its semiclassical far-field approximation. In the ballistic
semiclassical regime this gives the standard time-of-flight relation. As a
general arrival-time probability law, however, the current is problematic:
even states whose momentum support is entirely positive can exhibit quantum
backflow, for which the current at the detector becomes negative
\cite{BrackenMelloy1994}. The unrestricted Schr\"odinger current therefore
cannot itself be interpreted as a positive arrival-time probability density
and does not define an arrival-time POVM. These different constructions need
not yield identical detection statistics, and several direct comparisons have
exhibited experimentally distinguishable predictions
\cite{DasStruyve2021,DasNoth2021,DasEtAl2025,CavendishDas2025}.

A recent positive-operator formulation showed, through the MS--Kijowski
identity, that a normalized member of the generalized MS family yields the
same one-sided arrival-time POVM as Kijowski's construction
\cite{Marchewka2026}. This is an equality of the resulting POVMs, not an
identification of the physical processes represented by the two constructions.

The aim of the present paper is to compare the generalized MS detector
process with the absorbing Robin boundary. We show that the one-parameter
Robin absorption family is contained within the broader MS spectral-response
family, and we compare their absorption, reflection, and asymptotic behavior.

\section{The MS and ABC absorption currents}
\label{sec:MS_ABC_currents}

Let $\mathcal F_D$ and $\mathcal F_\kappa$ denote the Dirichlet and Robin
spectral transforms. For the same spatial initial state $\psi_0$,
\[
g_{\rm MS}=\mathcal F_D\psi_0,\qquad
g_{\rm ABC}=\mathcal F_\kappa\psi_0,\qquad
g_{\rm MS}\ne g_{\rm ABC}\quad\hbox{in general}.
\]
Both spectral coefficients are normalized independently:
\[
\int_0^\infty |g_{\rm MS}(k)|^2\,dk
=\int_0^\infty |g_{\rm ABC}(k)|^2\,dk=1,
\qquad
E_k=\frac{\hbar^2k^2}{2m}.
\]
When comparing the two detector laws below, $g(k)$ denotes a common normalized
one-sided incoming spectral amplitude. This statistical comparison should not be
confused with identifying $g_{\rm MS}$ and $g_{\rm ABC}$ for a fixed spatial
state $\psi_0$.

\subsection{The MS unidirectional current}

The MS construction can be described within the path-integral formalism.
Consider a free particle on the negative half-line $x<0$, with a detector at
the boundary $x=0$. The restricted state is constructed from paths that have
not reached the detector at any earlier time, and its infinitesimal free
propagation defines a finite positive one-sided current at the boundary. This
is the MS unidirectional first-arrival current
\cite{MarchewkaSchuss1998,MarchewkaSchuss2000,MarchewkaSchuss2002}.
The original hazard-based use of this current was compared with other
arrival-time proposals in Ref.~\cite{DasStruyve2021}; here we instead use its
generalized positive-operator formulation.

In the generalized spectral-response formulation, the current is
\begin{equation}
\boxed{
J_{\rm UD}^{(\lambda)}(t)
=
\frac{2\hbar}{m\pi^2}
\left|
\int_0^\infty dk\,
k\sqrt{\lambda(k)}\,g(k)e^{-iE_kt/\hbar}
\right|^2 .
}
\label{eq:MS_spectral_current}
\end{equation}
Here $g(k)$ specifies the one-sided spectral state, while the non-negative function
$\lambda(k)$ represents the spectral response of the particle--detector
interaction. Since the response enters at the amplitude level, the different
momentum components remain coherently combined in the time-dependent current.

As shown in Ref.~\cite{Marchewka2026}, this current defines a POVM-admissible
absorption law when $\lambda(k)$ satisfies the appropriate subnormalization
condition. The same condition follows directly from the integrated current,
\begin{equation}
P_\lambda[g]
=
\int_{-\infty}^{\infty}J_{\rm UD}^{(\lambda)}(t)\,dt
=
\frac{4}{\pi}\int_0^\infty dk\,k\lambda(k)|g(k)|^2 .
\label{eq:MS_integrated_current}
\end{equation}
Requiring $0\leq P_\lambda[g]\leq1$ for every normalized spectral state gives
\begin{equation}
\boxed{
0\leq\lambda(k)\leq\lambda_{\rm full}(k),
\qquad
\lambda_{\rm full}(k)=\frac{\pi}{4k},
}
\label{eq:MS_subnormalization}
\end{equation}
for almost every $k>0$. The upper response satisfies the
resolution-of-the-identity condition $P_{\lambda_{\rm full}}[g]=1$ for every
normalized $g(k)$.

It is convenient to define the dimensionless spectral absorption fraction
\begin{equation}
\eta_\lambda(k)
=
\frac{\lambda(k)}{\lambda_{\rm full}(k)}
=
\frac{4k\lambda(k)}{\pi},
\qquad
0\leq\eta_\lambda(k)\leq1,
\label{eq:MS_detection_fraction}
\end{equation}
so that
\[
P_\lambda[g]
=
\int_0^\infty dk\,\eta_\lambda(k)|g(k)|^2.
\]
The generalized MS family therefore includes both the fully normalized
response and subnormalized responses describing partial absorption.

\subsection{The absorbing Robin boundary}

The description of arrival by irreversible evolution and dissipative boundary
conditions was developed by Werner, who constructed arrival-time observables
from contraction semigroups describing absorption \cite{Werner1987}. The
absorbing-boundary rule and its interpretation as an ideal detection-time law
have subsequently been developed in a series of works by Tumulka and
collaborators
\cite{Tumulka2022,Tumulka2022Many,Tumulka2024Derivation,
GoldsteinTumulkaZanghi2024,GoldsteinTumulkaZanghiSpin2024,
JozaniTumulka2026}.

The ABC model describes a free particle on the half-line $x<0$,
\[
i\hbar\partial_t\psi(x,t)
=
-\frac{\hbar^2}{2m}\partial_x^2\psi(x,t),
\]
subject to the absorbing Robin boundary condition
\begin{equation}
\partial_x\psi(0,t)=i\kappa\psi(0,t),
\qquad \kappa>0.
\label{eq:ABC_Robin}
\end{equation}
This condition can be obtained as the limiting description of a thin absorbing
layer \cite{Tumulka2023}; here the limiting boundary law itself is taken as the
detector model.

The probability remaining in the exterior region is
\[
P_{\rm nonabs}^{\rm ABC}(t)
=
\int_{-\infty}^{0}|\psi(x,t)|^2\,dx .
\]
The continuity equation and Eq.~\eqref{eq:ABC_Robin} give
\[
-\frac{d}{dt}P_{\rm nonabs}^{\rm ABC}(t)=j_{\rm ABC}(0,t),
\]
where
\begin{equation}
\boxed{
j_{\rm ABC}(0,t)
=
\frac{\hbar\kappa}{m}|\psi(0,t)|^2\geq0.
}
\label{eq:ABC_boundary_current}
\end{equation}
The positive loss of norm therefore defines the ABC absorption-time density.

For a monochromatic incident component,
\[
\psi_k(x)=e^{ikx}+r_\kappa(k)e^{-ikx},
\qquad k>0,
\]
Eq.~\eqref{eq:ABC_Robin} gives
\[
r_\kappa(k)=\frac{k-\kappa}{k+\kappa}.
\]
The corresponding absorbed and reflected fractions are
\begin{equation}
A_\kappa(k)=\frac{4k\kappa}{(k+\kappa)^2},
\qquad
R_\kappa(k)=\left(\frac{k-\kappa}{k+\kappa}\right)^2,
\qquad
A_\kappa(k)+R_\kappa(k)=1.
\label{eq:ABC_absorption_reflection}
\end{equation}
Thus, a fixed Robin boundary generally describes partial absorption: the same
boundary condition determines both the absorbed and reflected fractions.
Perfect absorption occurs only for the monochromatic component $k=\kappa$.

For the Robin scattering realization, let $g(k)$ be the asymptotic
incoming spectral amplitude. The exterior state is
\[
\psi(x,t)
=
\frac{1}{\sqrt{2\pi}}
\int_0^\infty dk\,g(k)
\left[e^{ikx}+r_\kappa(k)e^{-ikx}\right]e^{-iE_kt/\hbar}.
\]
Since $1+r_\kappa(k)=2k/(k+\kappa)$, its boundary value is
\[
\psi(0,t)
=
\sqrt{\frac{2}{\pi}}
\int_0^\infty dk\,
\frac{k}{k+\kappa}\,g(k)e^{-iE_kt/\hbar}.
\]
Substitution into Eq.~\eqref{eq:ABC_boundary_current} gives
\begin{equation}
\boxed{
j_{\rm ABC}(0,t)
=
\frac{2\hbar\kappa}{\pi m}
\left|
\int_0^\infty dk\,
\frac{k}{k+\kappa}\,g(k)e^{-iE_kt/\hbar}
\right|^2 .
}
\label{eq:ABC_packet_current}
\end{equation}

The generalized MS family permits both complete and partial absorption through
the choice of $\lambda(k)$, whereas the fixed Robin model produces a particular
momentum-dependent partial-absorption law together with reflection. We now ask
whether this independently defined Robin law can be represented exactly by an
admissible generalized MS response.

\section{Matching the ABC and generalized MS absorption currents}

Equations~\eqref{eq:MS_spectral_current} and
\eqref{eq:ABC_packet_current} have the same coherent spectral structure. Their
equality for every admissible spectral coefficient and at every time is therefore obtained by
matching the spectral amplitudes,
\[
\frac{1}{\pi}k\sqrt{\lambda_\kappa(k)}
=
\sqrt{\frac{\kappa}{\pi}}\frac{k}{k+\kappa}.
\]
It follows that
\begin{equation}
\boxed{
\lambda_\kappa(k)=\frac{\pi\kappa}{(k+\kappa)^2}.
}
\label{eq:MS_Robin_response}
\end{equation}
Substitution into Eq.~\eqref{eq:MS_spectral_current} gives
\begin{equation}
\boxed{
J_{\rm UD}^{(\lambda_\kappa)}(t)=j_{\rm ABC}(0,t),
\qquad \forall t,
}
\label{eq:MS_ABC_time_identity}
\end{equation}
for every admissible spectral coefficient $g(k)$. The calibrated MS response
therefore reproduces the complete time-dependent ABC absorption current, not
merely its time integral.

Because the identity holds for the complete time-dependent current, integrating
from the remote past gives equal cumulative absorption, and therefore equal
nonabsorption probabilities, for the same incoming spectral amplitude:
\[
P_{\rm not\,abs}^{\rm MS}(t)=P_{\rm not\,abs}^{\rm ABC}(t),
\qquad \forall t.
\]
This is an equality of detector statistics, not an identification of the
underlying state dynamics.

The correspondence also holds momentum component by momentum component. From
Eqs.~\eqref{eq:MS_detection_fraction} and \eqref{eq:MS_Robin_response},
\begin{equation}
\boxed{
\eta_{\lambda_\kappa}(k)
=
\frac{4k\kappa}{(k+\kappa)^2}
=
A_\kappa(k),
\qquad k>0.
}
\label{eq:eta_equals_A}
\end{equation}
Consequently, the complementary spectral fraction is
$1-\eta_{\lambda_\kappa}(k)=R_\kappa(k)$. Because the calibration was obtained
at the amplitude level, it also preserves interference between distinct
momentum components. Equation~\eqref{eq:MS_ABC_time_identity} therefore holds
for every coherent spectral state, rather than only for individual momentum
components or for the integrated absorption probability.

The Robin family is therefore embedded in the generalized MS family: each value
of the single Robin parameter $\kappa$ selects one particular admissible
response function $\lambda_\kappa(k)$. The MS construction, however, allows
more general admissible functions $\lambda(k)$, including
$\lambda_{\rm full}(k)=\pi/(4k)$, which gives unit absorption efficiency for
every wave number and cannot be generated by any fixed $\kappa$.

\section{Ballistic fixed-ray far-field characterization of the spectral response}
\label{sec:far_field_response}

Time-of-flight measurements routinely infer particle velocity and momentum
from detection times. Their interpretation commonly relies on the far-field
approximation, in which the particle propagates freely between preparation and
detection. This suggests that the recovery of the standard free-particle behavior
may serve as a useful consistency criterion for a refined time-of-arrival
description. 

We consider the limit $L,t\to\infty$ with $L/t$ fixed, where $L$ is the
propagation distance and $t$ the detection time. For a one-sided freely
propagating packet,
\[
\psi_{\rm free}(L,t)
=
\frac{1}{\sqrt{2\pi}}
\int_0^\infty dk\,g(k)e^{i\Phi(k)},
\qquad
\Phi(k)=kL-\frac{\hbar k^2}{2m}t,
\]
and
\[
j_{\rm free}(L,t)
=
\frac{\hbar}{m}\operatorname{Im}
\left[\psi_{\rm free}^*(L,t)\partial_L\psi_{\rm free}(L,t)\right].
\]
At the same detector position, the generalized MS current is
\[
J_{\rm UD}^{(\lambda)}(L,t)
=
\frac{2\hbar}{m\pi^2}
\left|
\int_0^\infty dk\,k\sqrt{\lambda(k)}\,g(k)e^{i\Phi(k)}
\right|^2 .
\]

Under the standard regularity assumptions, stationary phase selects
\[
\Phi'(k_*)=0,
\qquad
k_*=\frac{mL}{\hbar t},
\qquad
\frac{\hbar k_*}{m}=\frac{L}{t}.
\]
Thus, each far-field ray selects the momentum component whose free velocity
reaches the detector at time $t$. The leading asymptotic currents are
\begin{align}
j_{\rm free}(L,t)
&\sim \frac{k_*}{t}|g(k_*)|^2,
\label{eq:free_far_field_current}\\
J_{\rm UD}^{(\lambda)}(L,t)
&\sim
\frac{4k_*\lambda(k_*)}{\pi}\,
\frac{k_*}{t}|g(k_*)|^2,
\label{eq:MS_far_field_current}\\
j_{\rm ABC}(L,t)
&\sim
A_\kappa(k_*)\,
\frac{k_*}{t}|g(k_*)|^2.
\label{eq:ABC_far_field_current}
\end{align}
Here, in the third line, the Robin boundary is understood to be translated to
$x=L$. Thus the stationary-phase time-of-flight relation itself remains
unchanged: $k_*=mL/(\hbar t)$. However, the detector response reweights each
far-field ray by the momentum-dependent efficiency $\eta_\lambda(k_*)$.
Since different arrival times correspond to different values of $k_*$, a
momentum-dependent response changes their relative weights and therefore
distorts the observed time-of-flight distribution; it is not, in general,
equivalent to an overall loss of detection efficiency. For the fully
normalized MS response,
\[
\lambda_{\rm full}(k)=\frac{\pi}{4k}
\quad\Longrightarrow\quad
J_{\rm UD}^{(\lambda_{\rm full})}(L,t)\sim j_{\rm free}(L,t).
\]
Thus the fully normalized MS calibration preserves the free asymptotic
current ray by ray. For the Robin calibration, the exact identity
$j_{\rm ABC}=J_{\rm UD}^{(\lambda_\kappa)}$ gives the third line, with
$A_\kappa(k)=4k\kappa/(k+\kappa)^2\leq1$. The factor $A_\kappa(k)$ equals one
only at $k=\kappa$ and varies with momentum. The Robin response therefore
changes the relative momentum weights and distorts the far-field time-of-flight
signal relative to the Schrödinger current, rather than merely reducing it by
an overall constant. This agrees with Ref.~\cite{CavendishDas2025}.

Requiring the efficiency to be constant for every ray gives
\[
\frac{4k\lambda(k)}{\pi}=c,
\qquad 0<c\leq1,
\]
and hence
\begin{equation}
\boxed{
\lambda(k)=c\,\frac{\pi}{4k}.
}
\label{eq:far_field_response_family}
\end{equation}
The constant $c$ is a momentum-independent absorption efficiency. Exact
agreement with the absolute free current corresponds to $c=1$, whereas
$0<c<1$ preserves the far-field shape with an overall reduced detection
probability.

The fixed-ray limit used above should be distinguished from the
fixed-distance late-time limit, in which $L$ is held fixed and $t\to\infty$.
Then $k_*=mL/(\hbar t)$ approaches the endpoint $k=0$, and the asymptotics are
controlled by the low-momentum behavior of the state. In particular, for the
localized Dirichlet preparations for which $g(k)=O(k)$ as $k\to0$, the
Robin-calibrated amplitude behaves as $k\sqrt{\lambda_\kappa(k)}g(k)=O(k^2)$,
and therefore
\[
J_{\rm UD}^{(\lambda_\kappa)}(L,t)=O(t^{-3})\qquad (L\ \text{fixed},\ t\to\infty).
\]
This late-time endpoint asymptotic belongs to a different asymptotic regime
from the ballistic fixed-ray result above; the distinction has also been
emphasized in Ref.~\cite{RafsanjaniCavendish2026}.

\section*{Discussion and conclusions}

Within detector-based approaches to quantum arrival time, the detection event
is identified with the arrival of the particle at the detector. In this
setting, we have established an exact correspondence between the absorbing
Robin boundary condition and the generalized MS unidirectional-current family.
The calibrated response \eqref{eq:MS_Robin_response} reproduces the complete
ABC absorption current for every admissible one-sided spectral amplitude $g$ and at every time. Since
the matching holds at the amplitude level, it also preserves interference
between different momentum components.

The correspondence extends to the complementary outcomes. The MS absorption
fraction coincides, momentum by momentum, with the Robin absorption coefficient,
while its complementary fraction coincides with the Robin reflection
coefficient. The generalized MS family therefore contains the ABC absorption
and reflection statistics as a particular calibrated case and, in this
statistical sense, generalizes the one-parameter Robin family. The
correspondence is statistical and does not identify the MS and ABC boundary
dynamics.

We also examined the 
ballistic fixed-ray far-field behavior of the absorption currents. In
agreement with Ref.~\cite{CavendishDas2025}, a fixed Robin response does not
reproduce the universal 
ballistic fixed-ray free-particle asymptotics. More generally,
Eq.~\eqref{eq:far_field_response_family} shows that preservation of the free
far-field form requires $\lambda(k)=c\pi/(4k)$. Exact agreement and full POVM
normalization select $c=1$, while $0<c<1$ describes partial absorption without
changing the far-field shape. The Robin-calibrated law and the class preserving

ballistic fixed-ray free far-field propagation are therefore distinct detector calibrations within
the generalized MS family. The exact result established here is confined
to the free scalar one-dimensional single-pass half-line setting; extensions
with repeated encounters, additional channels, or spin require further
structure.

\begingroup
\small
\bibliographystyle{unsrt}
\bibliography{references}

@article{Kijowski1974,
  author  = {Kijowski, Jerzy},
  title   = {On the Time Operator in Quantum Mechanics and the Heisenberg Uncertainty Relation for Energy and Time},
  journal = {Reports on Mathematical Physics},
  volume  = {6},
  number  = {3},
  pages   = {361--386},
  year    = {1974},
  doi     = {10.1016/S0034-4877(74)80004-2}
}

@article{DasStruyve2021,
  author  = {Das, Siddhant and Struyve, Ward},
  title   = {Questioning the Adequacy of Certain Quantum Arrival-Time Distributions},
  journal = {Physical Review A},
  volume  = {104},
  pages   = {042214},
  year    = {2021},
  doi     = {10.1103/PhysRevA.104.042214}
}

@article{DasNoth2021,
  author  = {Das, Siddhant and N{\"o}th, Markus},
  title   = {Times of Arrival and Gauge Invariance},
  journal = {Proceedings of the Royal Society A},
  volume  = {477},
  number  = {2250},
  pages   = {20210101},
  year    = {2021},
  doi     = {10.1098/rspa.2021.0101}
}

@article{DasEtAl2025,
  author  = {Das, Siddhant and Deckert, Dirk-Andr{\'e} and Kellers, Leopold and Krekels, Simon and Struyve, Ward},
  title   = {Double-Slit Experiment Revisited},
  journal = {Annals of Physics},
  volume  = {479},
  pages   = {170054},
  year    = {2025},
  doi     = {10.1016/j.aop.2025.170054}
}

@article{MarchewkaSchuss1998,
  author  = {Marchewka, A. and Schuss, Z.},
  title   = {Feynman Integrals with Absorbing Boundaries},
  journal = {Physics Letters A},
  volume  = {240},
  number  = {4--5},
  pages   = {177--184},
  year    = {1998}
}

@article{MarchewkaSchuss2000,
  author  = {Marchewka, A. and Schuss, Z.},
  title   = {Path-Integral Approach to the Schr\"odinger Current},
  journal = {Physical Review A},
  volume  = {61},
  pages   = {052107},
  year    = {2000}
}

@article{MarchewkaSchuss2002,
  author  = {Marchewka, A. and Schuss, Z.},
  title   = {Measurement as Absorption of Feynman Trajectories: Collapse of the Wave Function Can Be Avoided},
  journal = {Physical Review A},
  volume  = {65},
  pages   = {042112},
  year    = {2002}
}

@article{Tumulka2023,
  author  = {Tumulka, Roderich},
  title   = {Absorbing Boundary Condition as Limiting Case of Imaginary Potentials},
  journal = {Communications in Theoretical Physics},
  volume  = {75},
  number  = {1},
  pages   = {015103},
  year    = {2023},
  doi     = {10.1088/1572-9494/ac9bea}
}

@article{Werner1987,
  author  = {Werner, Reinhard},
  title   = {Arrival Time Observables in Quantum Mechanics},
  journal = {Annales de l'Institut Henri Poincar{\'e}, Physique Th{\'e}orique},
  volume  = {47},
  number  = {4},
  pages   = {429--449},
  year    = {1987},
  url     = {https://www.numdam.org/item/AIHPA_1987__47_4_429_0/}
}

@article{Tumulka2022,
  author  = {Tumulka, Roderich},
  title   = {Distribution of the Time at Which an Ideal Detector Clicks},
  journal = {Annals of Physics},
  volume  = {442},
  pages   = {168910},
  year    = {2022},
  doi     = {10.1016/j.aop.2022.168910}
}

@article{Tumulka2022Many,
  author  = {Tumulka, Roderich},
  title   = {Detection-Time Distribution for Several Quantum Particles},
  journal = {Physical Review A},
  volume  = {106},
  pages   = {042220},
  year    = {2022},
  doi     = {10.1103/PhysRevA.106.042220}
}

@article{Tumulka2024Derivation,
  author  = {Tumulka, Roderich},
  title   = {On a Derivation of the Absorbing Boundary Rule},
  journal = {Physics Letters A},
  volume  = {494},
  pages   = {129286},
  year    = {2024},
  doi     = {10.1016/j.physleta.2023.129286}
}

@article{GoldsteinTumulkaZanghi2024,
  author  = {Goldstein, Sheldon and Tumulka, Roderich and Zangh{\`i}, Nino},
  title   = {Arrival Times Versus Detection Times},
  journal = {Foundations of Physics},
  volume  = {54},
  number  = {5},
  pages   = {63},
  year    = {2024},
  doi     = {10.1007/s10701-024-00798-y}
}

@article{GoldsteinTumulkaZanghiSpin2024,
  author  = {Goldstein, Sheldon and Tumulka, Roderich and Zangh{\`i}, Nino},
  title   = {On the Spin Dependence of Detection Times and the Nonmeasurability of Arrival Times},
  journal = {Scientific Reports},
  volume  = {14},
  pages   = {3775},
  year    = {2024},
  doi     = {10.1038/s41598-024-53777-8}
}

@article{CavendishDas2025,
  author  = {Cavendish, Will and Das, Siddhant},
  title   = {Absorbing Detectors versus Scattering Theory},
  journal = {Physical Review A},
  volume  = {112},
  pages   = {062217},
  year    = {2025},
  doi     = {10.1103/332p-4nl1}
}

@misc{Marchewka2026,
  author        = {Marchewka, Avi},
  title         = {The Unidirectional Current as First Arrival-Time POVM: An MS--Kijowski Identity, Physical Interpretation, and Mathematical Applications},
  year          = {2026},
  eprint        = {2608.16510},
  archivePrefix = {arXiv},
  primaryClass  = {quant-ph},
  note          = {arXiv:2608.16510 [quant-ph]}
}

@article{JozaniTumulka2026,
  author  = {Jozani, Alireza and Tumulka, Roderich},
  title   = {{Detection Time Distribution Predicted Using Absorbing Boundary Conditions and Imaginary Potentials}},
  journal = {Physical Review Research},
  volume  = {8},
  pages   = {033072},
  year    = {2026},
  doi     = {10.1103/g4wn-ctbl}
}

@article{BrackenMelloy1994,
  author  = {Bracken, A. J. and Melloy, G. F.},
  title   = {Probability Backflow and a New Dimensionless Quantum Number},
  journal = {Journal of Physics A: Mathematical and General},
  volume  = {27},
  number  = {6},
  pages   = {2197--2211},
  year    = {1994},
  doi     = {10.1088/0305-4470/27/6/040}
}

@misc{RafsanjaniCavendish2026,
  author        = {Rafsanjani, Ali Ayatollah and Cavendish, Will},
  title         = {The Arrival Position Problem in Quantum Mechanics},
  year          = {2026},
  eprint        = {2607.03538},
  archivePrefix = {arXiv},
  primaryClass  = {quant-ph},
  note          = {arXiv:2607.03538 [quant-ph]}
}
\endgroup

\end{document}